\documentclass[12pt]{iopart}

\usepackage{natbib}
\usepackage{graphicx}
\usepackage{color} 
\usepackage{hyperref}

\begin{document}

\title{Validity of models for Dreicer generation of runaway electrons in dynamic scenarios}

\author{S. Olasz$^{1,2}$, O. Embreus$^3$, M. Hoppe$^3$, M. Aradi$^4$, D. Por$^1$, T. Jonsson$^5$, D. Yadikin$^6$, G.I. Pokol$^{1,2}$, EU-IM Team$^*$}

\address{$^1$ Institute of Nuclear Techniques, Budapest University of Technology and Economics,
Műegyetem rkp. 3, Budapest 1111, Hungary}
\address{$^2$ Fusion Plasma Physics Department, Centre For Energy Research, Budapest, Hungary}
\address{$^3$ Department of Physics, Chalmers University of
Technology,SE-41296 Göteborg, Sweden}
\address{$^4$ Barcelona Supercomputing Center (BSC), Barcelona, Spain}
\address{$^5$ KTH Royal Institute of Technology, Stockholm, Sweden}
\address{$^6$ Department of Space, Earth and Environment, Chalmers University of
Technology,SE-41296 Göteborg, Sweden}
\address{$^*$ https://users.euro-fusion.org/eu-im}
\ead{olaszsoma@reak.bme.hu}

\begin{indented}
\item[]December 2020
\end{indented}
\begin{abstract}
Runaway electron modelling efforts are motivated by the risk these energetic particles pose to large fusion devices. The sophisticated kinetic models can capture most features of the runaway electron generation but have high computational costs, which can be avoided by using computationally cheaper reduced kinetic codes. This paper compares the reduced kinetic and kinetic models to determine when the former solvers, based on analytical calculations assuming quasi--stationarity, can be used. The Dreicer generation rate is calculated by two different solvers in parallel in a workflow developed in the European Integrated Modelling framework, and this is complemented by calculations of a third code that is not yet integrated into the framework. \textsc{Runaway Fluid}, a reduced kinetic code, \textsc{NORSE}, a kinetic code using non-linear collision operator, and \textsc{DREAM}, a linearized Fokker--Planck solver, are used to investigate the effect of a dynamic change in the electric field for different plasma scenarios spanning across the whole tokamak--relevant range. We find that on time scales shorter than or comparable to the electron--electron collision time at the critical velocity for runaway electron generation, kinetic effects not captured by reduced kinetic models play an important role. This characteristic time scale is easy to calculate and can reliably be used to determine whether there is a need for kinetic modelling or cheaper reduced kinetic codes are expected to deliver sufficiently accurate results. This criterion can be automated, and thus it can be of great benefit for the comprehensive self--consistent modelling frameworks that are attempting to simulate complex events such as tokamak start--up or disruptions.
\end{abstract}

\section{Introduction}
Correct understanding and accurate simulations of runaway electron generation are of great importance as they pose a severe risk to the upcoming ITER experiment~\citep{Boozer17, Lehnen15}. Runaway electrons can appear when large electric fields are generated in various events, such as disruptions or the start--up of a tokamak discharge~\citep{Zeng17, Sharma88, deVries20}. A large fraction of the pre--disruption current can be converted into runaway current on a few milliseconds timescale, and on large devices such as JET, the runaway current can reach the magnitudes of megaamps~\citep{Plyusnin06, Plyusnin17}.

Runaway electrons can be generated in plasmas because of the reduced friction force experienced by fast particles compared to the thermal population. An applied electric field can create a net accelerating region, called runaway region, in momentum space where the particles are accelerated to high energies until synchrotron radiation losses can balance the acceleration of the electric field. A primary runaway electron generation mechanism, named Dreicer generation, is the diffusion of particles in velocity space into the runaway region due to collisions~\citep{Dreicer60}. The expression was generalized to the relativistic case by J. Connor and R. Hastie~\citep{Connor75}. Both formulas describe the generation in steady--state cases and are used in the present paper as a basis for comparison. 

The steady--state generation rates are often used in numerical models to calculate the runaway electron generation -- these models are often referred to as fluid or reduced kinetic codes. One of these codes is \textsc{Runaway Fluid}\footnote{https://github.com/osrep/Runafluid}, which is used in the European Integrated Modeling (EU-IM) framework~\citep{Pokol19}, and is also integrated into the ITER Analysis and Modelling Suite (IMAS)~\citep{Imbeaux15}. \textsc{Runaway Fluid} uses the analytical formulas by Connor and Hastie~\citep{Connor75} for the Dreicer generation rate to estimate the runaway electron density. The European Transport Simulator workflow with the \textsc{Runaway Fluid} has been benchmarked against the \textsc{GO} code~\citep{Pokol19}. \textsc{GO}~\citep{Papp13} is a transport solver for cylindrical geometry for the study of runaway electrons in disruptions which has a similar implementation of generation rates as the \textsc{Runaway Fluid} code. Reduced kinetic models have also been used in other comprehensive simulation tools, such as JOREK~\citep{Bandaru19} and ASTRA-STRAHL~\citep{Linder20}. The common usage of these models in integrated modelling tools motivates the study of the range of applicability of this approach.

Kinetic solvers give a wider range of opportunities in runaway modelling. The NORSE code~\citep{Stahl16}\footnote{{https://github.com/hoppe93/NORSE}} solves the kinetic equation with a non-linear collision operator. Therefore, it is suitable for slide--away scenarios, as it takes runaway--runaway interaction into account. The \textsc{DREAM}~\citep{DREAM}\footnote{https://github.com/chalmersplasmatheory/DREAM} code, on the other hand, solves the bounce averaged Fokker--Planck equation with a linearized collision operator, while analytically solving the background plasma evolution in a manner similar to \textsc{GO}. There are several other kinetic solvers commonly used for the study of runaway electrons, including, \textsc{CODE}~\citep{Landreman13, Stahl2016}, \textsc{KORC}~\citep{del-Castillo-Negrete18, Beidler20} and \textsc{LUKE}~\citep{Peysson14, Decker16}, all considering the problem to different degrees of complexity. The present paper shows results from \textsc{NORSE} and \textsc{DREAM}.

These kinetic solvers are becoming more and more accurate in simulating the evolution of the runaway electron distribution functions, but the price of the accuracy is paid in computation time. Therefore, self--consistent simulations tend to use fluid--like models, which provide satisfying results in most cases, but the underlying steady--state assumptions fail in highly dynamic scenarios. In the present paper, we study the time evolution of the primary generation rate with various models in scenarios with an instantaneously introduced electric field. This approach is inspired by the use of the step function for characterizing the system response to rapid changes in system theory~\citep{Feucht90}. For this purpose, \textsc{Runaway Fluid} and \textsc{NORSE} are integrated into a workflow called the Runaway Electron Test Workflow in the EU-IM framework~\citep{Pokol19, Falchetto14}. The workflow is used to study the effect of the electric field on the time evolution of the primary runaway generation for different electron densities and temperatures with the main focus on specifying the conditions in which it is safe to use the fluid-like approach. A case of rapidly evolving temperature, referred to as hot-tail generation, is not addressed, as it has been considered in other papers recently~\citep{Smith08, Bjork20, Aleynikov17}.

Four cases with differing electron temperature and density values have been selected for the current study in a way to span the whole tokamak-relevant range of plasma parameters. Consequently, the chosen cases are relevant to extreme cases of plasma operation where runaway electron generation might be expected. Our study focuses on the range of applicability of the analytical formulas and the linear collision operator for these cases. The Dreicer generation rate and runaway electron density are calculated. We find that a kinetic effect can lead to a peak in the Dreicer generation rate on short timescales. The time over which this peak occurs can be characterized by the electron-electron collision time at the critical velocity for runaway electron generation. The modelling tool used for the simulations is described in Section~\ref{sec:model}, the results for the various cases are presented in Section~\ref{sec:results}. The other primary generation rates, such as hot-tail generation~\citep{Helander04}, and the secondary or avalanche generation~\citep{Rosenbluth97} is not considered in the present paper but their significance on our results is elaborated in Section~\ref{sec:discussion}, where we also motivate the usage of the step function for the change of the electric field. Finally, the paper is concluded in Section~\ref{sec:conclusion}.

\section{Model description}\label{sec:model}

The present work is carried out in the EUROfusion Integrated Modeling framework (EU-IM), which was developed to enable the coupling of different codes simulating different physical phenomena in complex workflows. This is achieved by defining a standardized data structure and providing standardized access to the named data structure so that the different codes can exchange input and output data with ease~\citep{Falchetto14}. The framework also allows for experimental data to be imported into the data structure. The integration of physics codes into workflows is managed by the Kepler graphical workflow engine \footnote{https://kepler-project.org}. The \textsc{European Transport Simulator (ETS)} workflow aims to simulate tokamak discharges and has been proposed for self--consistent simulation of tokamak disruptions~\citep{Pokol19}. Each module in ETS used for simulating different physical aspects of discharges have multiple codes integrated, and it can be chosen which of these codes is used during a simulation.

The purpose of \textsc{Runaway Indicator}, the simplest runaway model in ETS, is to indicate when the physical parameters during a simulation are suitable for runaway electron generation. It calculates the critical electric field~\citep{Connor75} for given plasma parameters and indicates when an electric field larger than the critical field is present. The formula for the critical electric field used by Runaway Indicator and also Runaway Fluid is the one introduced by Connor and Hastie~\citep{Connor75},
\begin{equation}
E_c=\frac{n_e e^3 \ln\Lambda}{4 \pi \varepsilon_0^2 {m_e}c^2},
\label{Ec}
\end{equation}
where $m_e$ is the mass of the electron, $e$ is the elementary charge, $n_e$ is the electron density and $\ln\Lambda$ is the Coulomb logarithm~\citep{Huba13}. It also estimates the primary runaway electron generation rate with a simple formula and gives a warning if it is larger than a preset value~\citep{Pokol19}\footnote{{https://github.com/osrep}}.

During a simulation, \textsc{Runaway Fluid} can give a conservative estimate of the runaway electron population. It calculates the primary and secondary (avalanche) generation of runaway electrons. The analytical formulas derived by J. Connor and R. Hastie \cite{Connor75} are implemented for primary generation calculation, including the most general formula (63) in the paper and it was used for the simulations presented in the current paper. The avalanche generation rate is calculated using the formula derived by M.N. Rosenbluth and S.V. Putvinski~\citep{Rosenbluth97}. Both generation mechanisms can be augmented by the use of a number of different correction factors. Using these generation rates, \textsc{Runaway Fluid} estimates the runaway electron density and current \cite{Pokol19}. In the present work, no correction factor is used and the secondary generation is not calculated as we are focusing on the behavior of Dreicer generation for various plasma scenarios.

\textsc{NORSE} is a more advanced tool that solves the kinetic equation to calculate the electron distribution function~\citep{Stahl16}. It uses a non-linear collision operator, which makes it suitable for taking runaway-runaway interaction into account. It is also capable of simulating slide-away scenarios. The \textsc{NORSE} code is written in Matlab~\citep{Stahl16}. The interfacing of the code to the EU-IM framework was done using Python \footnote{{https://github.com/osrep}}. A Matlab Engine is used for the communication between the different programming languages, which allows for Python codes to execute Matlab commands. The Python interface is, in turn, integrated into the Kepler workflow as an actor, and it provides the input to the \textsc{NORSE} code and writes the output to the required data structure. The default \textsc{NORSE} output contains the electron distribution function, the corresponding momentum grid, the runaway density and current. Additional information, such as plasma parameters for the simulation, the value of the critical field, runaway electron density, current and generation rate is output as an option from \textsc{NORSE} in HDF5 format.

\begin{figure*}[t]
\begin{center}
\includegraphics[width=1\linewidth]{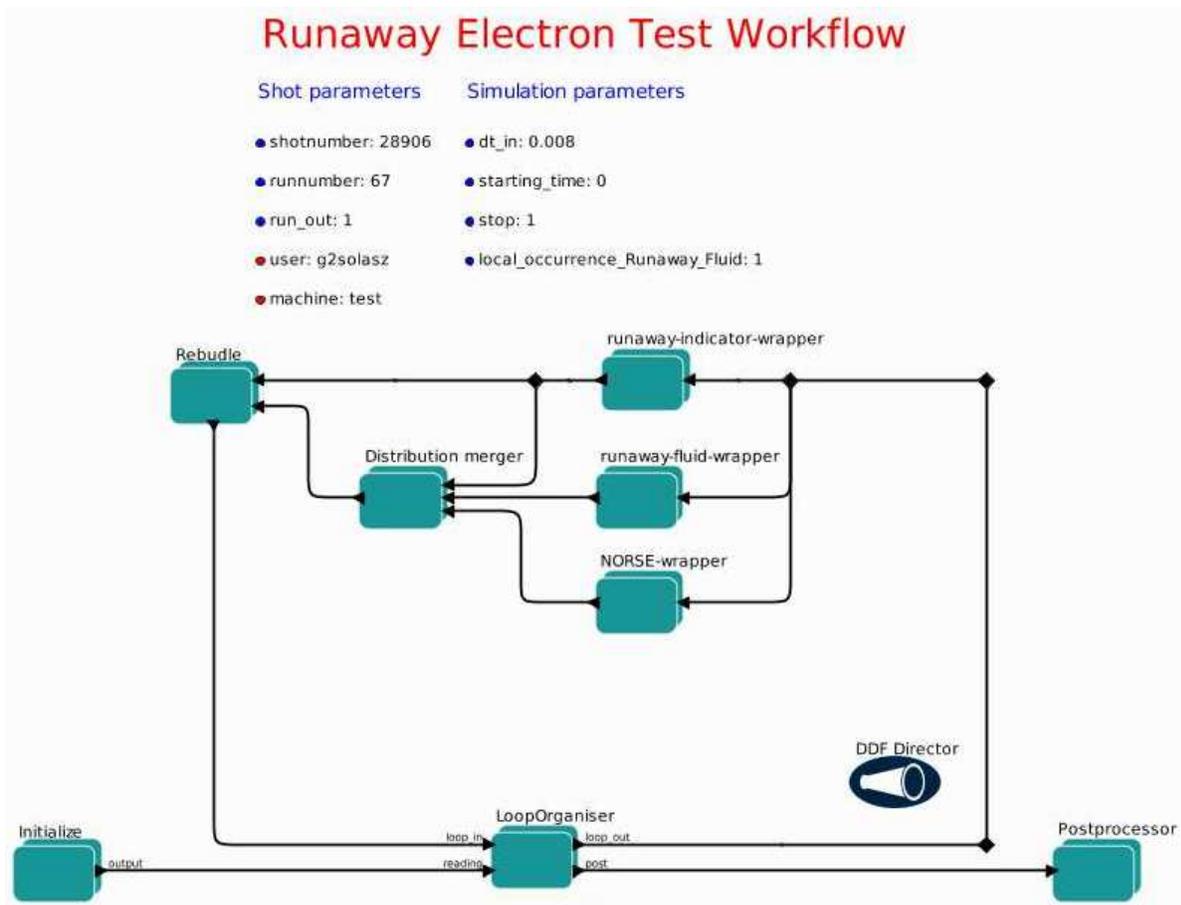}
\caption{The dedicated runaway electron workflow in the EU-IM framework. The boxes contain the models used in the workflow, and the lines connecting them represent the data bundle passing around the loop. The DDF (Dynamic dataflow) director is responsible for the control of the workflow execution.}
\label{Workflow}
\end{center}
\end{figure*}

These codes have been interfaced in the EU-IM framework to create so-called actors in Kepler workflows. The dedicated runaway electron workflow, the Runaway Electron Test Workflow~\footnote{{https://gforge-next.eufus.eu/svn/keplerworkflows} r2060} is shown in Figure~\ref{Workflow}. The boxes in the Figure are composite actors consisting of an embedded workflow themselves, which perform different actions during simulations. The black diamonds visible on the lines connecting the actors are used to organize the workflow layout and split the data bundle. The input parameters can be given under the shot parameters. The \texttt{shotnumber}, \texttt{runnumber} and \texttt{machine} parameters determine which data will be used from the database, while the run\textunderscore out parameter determines what run number the output will be stored in. Simulation parameters such as \texttt{time step}, \texttt{starting time}, and \texttt{number of iterations} can also be given. As \textsc{Runaway Fluid} and \textsc{NORSE} output are of the same data type, they have to be differentiated by a parameter called occurrence number. The \texttt{local\textunderscore occurrence\textunderscore Runaway\textunderscore fluid} is used for this purpose for \textsc{Runaway Fluid}, while the occurrence number for \textsc{NORSE} can be given at the actor level.

\begin{figure*}[ht]
\begin{center}
\includegraphics[width=1\linewidth]{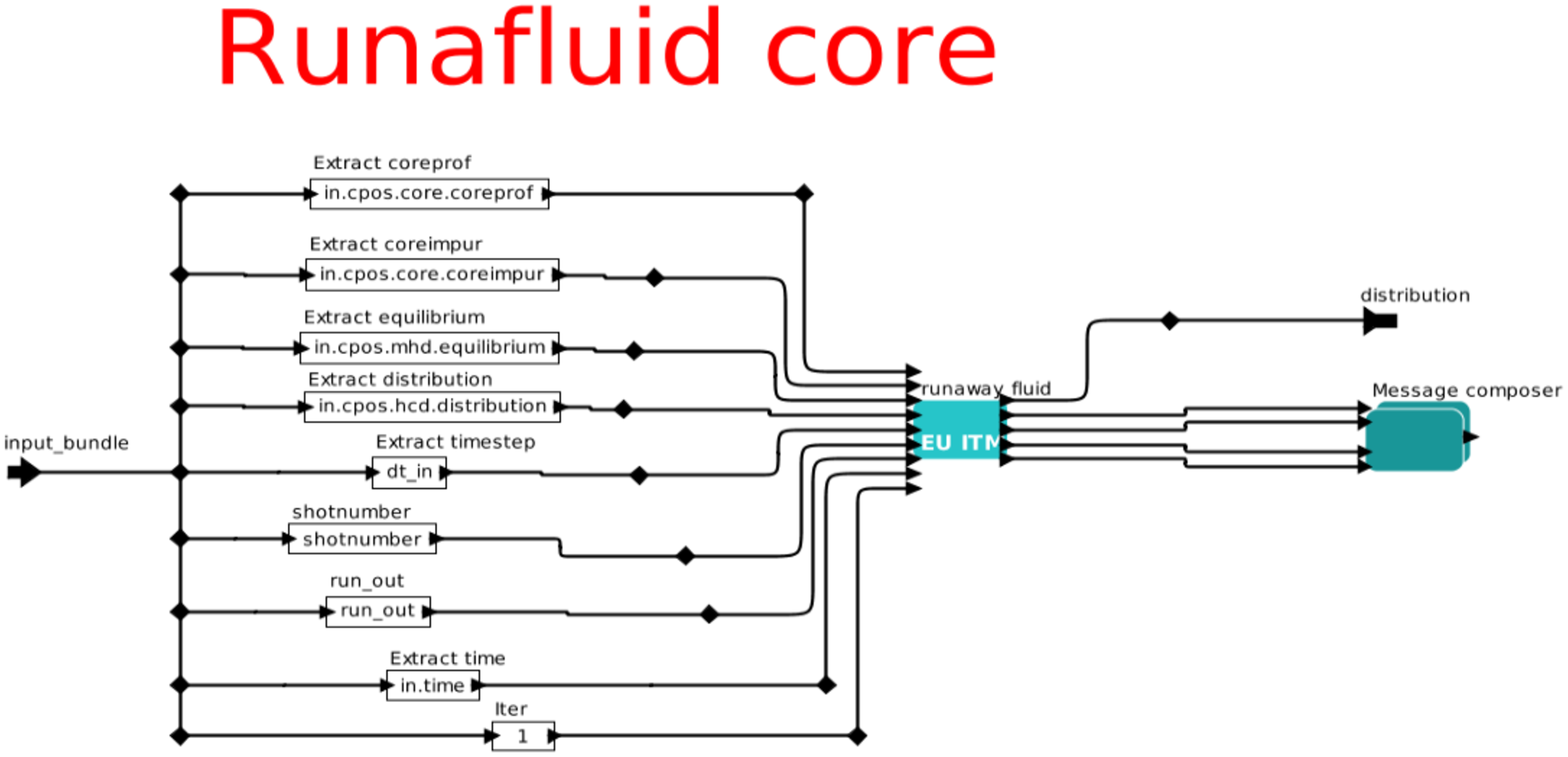}
\caption{The content of the runaway-fluid-wrapper composite actor. First the data bundle is processed and the necessary data is extracted. This is given to the Runaway Fluid code, which outputs the calculated data and possible run--time messages.}
\label{Rf}
\end{center}
\end{figure*}

The workflow starts at the \texttt{Initialize} composite actor, see Figure~\ref{Workflow}, where the input data is read from the database as specified by the simulation parameter settings. The main part of the workflow is the time loop, where the different physics codes are integrated. This is controlled by the \texttt{LoopOrganizer} which stops the loop after a preset number of iterations. The post-processing of the simulation results is done in the \texttt{Postprocessor} actor. The wrapper composite actors contain the physics codes, as shown for the \textsc{Runaway Fluid} in Figure~\ref{Rf}. First, the input data for the \texttt{Runaway Fluid} actor is extracted from the bundle and delivered to the code. The physical quantities are stored in dedicated data structures, such as \texttt{coreprof}. In these data structures, the required physical quantities are delivered to the model; \texttt{coreprof}, for examples, contains the plasma profiles, such as electron density and temperature profiles. These are used by \texttt{Runaway Fluid} to calculate the runaway electron density and current, which are output to a \texttt{distribution} data structure. Any possible run--time messages are output by the \texttt{Message composer} composite actor. The output is then passed onto the next workflow level, Figure~\ref{Workflow}, where the data from the different physical actors is merged in the \texttt{Distribution merger} composite actor. The \texttt{Rebundle} box merges this with the data bundle going around in the loop. The main advantage of this method is that the different codes can be run parallel with identical input and their output can be easily compared.

The \textsc{DREAM} kinetic solver has not yet been integrated into the EU-IM framework, as some parts of it are still under development \cite{DREAM}\footnote{https://github.com/chalmersplasmatheory/DREAM}. However, we use it here for studying the Dreicer generation rate obtained with a linearized test-particle collision operator. The \textsc{DREAM} code is specifically designed for the self--consistent study of runaway generation during tokamak disruptions and to solve a bounce-averaged Fokker--Planck equation, along with equations for background plasma parameters such as the electric field, electron temperature and ion densities. In this work, our main pieces of interest are transient effects on the Dreicer runaway generation rate and therefore we use the code to only solve for the electron distribution function in a homogeneous plasma with prescribed parameters, i.e.\ equivalent to \textsc{NORSE} but with a linearized test-particle collision operator.

\section{Simulation results}\label{sec:results}

In this section, the simulation results are presented. Firstly, the physical parameters used for the different scenarios are shown. The evolution of the distribution function from \textsc{NORSE} calculations is presented, and the calculated generation rate from the different models is given. The evolution of the generation rate based on the distribution function and its effects on the runaway electron density are discussed. Lastly, the timescales characterizing the kinetic effects are introduced.

\subsection{Evolution of distribution function}\label{sec:distribution}

\begin{table}[t]
\centering
\caption{The electrons density and temperature values for the various scenarios studied.}
\begin{tabular}{|l|l|l|l|l|}
\hline
& \multicolumn{1}{c|}{\begin{tabular}[c]{@{}c@{}}Low density\\ discharge (LD)\end{tabular}} 
& \multicolumn{1}{c|}{\begin{tabular}[c]{@{}c@{}}Start--up\\ phase (SU)\end{tabular}} 
& \multicolumn{1}{c|}{\begin{tabular}[c]{@{}c@{}}End of\\ disruption (ED)\end{tabular}} 
& \multicolumn{1}{c|}{\begin{tabular}[c]{@{}c@{}}Start of \\ disruption (SD)\end{tabular}} 
\\ \hline
Density {[}$\rm m^{-3}${]} & $5\cdot 10^{17}$ & $5\cdot 10^{17}$ & $10^{20}$ & $10^{20}$  \\ \hline
Temperature {[}eV{]} & 10000 & 300 & 300 & 10000 \\ \hline
\end{tabular}
\label{cases}
\end{table}

\begin{table}[b]
\centering
\caption{The various plasma parameters used for the simulations. The collision time at the critical velocity was found to be the relevant time scale.}
\begin{tabular}{|l|l|l|l|l|}
\hline
& \multicolumn{1}{c|}{\begin{tabular}[c]{@{}c@{}}Low density\\ discharge (LD)\end{tabular}}
& \multicolumn{1}{c|}{\begin{tabular}[c]{@{}c@{}}Start--up\\ phase (SU)\end{tabular}} 
& \multicolumn{1}{c|}{\begin{tabular}[c]{@{}c@{}}End of\\ disruption (ED)\end{tabular}} 
& \multicolumn{1}{c|}{\begin{tabular}[c]{@{}c@{}}Start of \\ disruption (SD)\end{tabular}} \\ \hline
Electric field {[}V/m{]} & $2.81 \cdot 10^{-3}$ & $2.96 \cdot 10^{-2}$ & 3.66 & $4.38 \cdot 10^{-1}$ \\ \hline
Critical field {[}V/m{]} & $5.06 \cdot 10^{-4}$ & $4.16 \cdot 10^{-4}$ & $6.98 \cdot 10^{-2}$ & $8.77 \cdot 10^{-2}$ \\ \hline
\begin{tabular}[c]{@{}l@{}}Normalized electric \\ field {[}-{]}\end{tabular} & 5.55 & 71 & 52.5 & 5 \\ \hline
\begin{tabular}[c]{@{}l@{}}Coulomb\\ logarithm {[}-{]}\end{tabular} & 19.9 & 16.3 & 13.7 & 17.2 \\ \hline
\begin{tabular}[c]{@{}l@{}}Collision time at\\ critical velocity {[}s{]}\end{tabular} & $2.58 \cdot 10^{-1}$ & $6.84 \cdot 10^{-3}$ & $6.42 \cdot 10^{-5}$ & $1.74 \cdot 10^{-3}$ \\ \hline
\end{tabular}
\label{params}
\end{table}

The simulations were carried out for the four different combinations of density and temperature values given in Table~\ref{cases} using the Runaway Electron Test Workflow. The different values were chosen to represent extreme cases of electron densities and temperature still relevant for tokamak operations. A low density and low temperature scenario is relevant for the start--up (SU) phase of tokamak discharges, where large electric fields are induced, which are capable of creating a runaway electron population~\citep{Sharma88, deVries20}. During the flat--top phase of a plasma discharge, a starting disruption (SD) might create a large enough electric field to generate runaway electrons despite high electron density initially at high temperature, while the low temperature and high density case is relevant for the end of a disruption (ED)~\citep{Papp13}. Finally, the high temperature low density case is relevant for low density discharges (LD) where runaway electron generation is studied~\citep{Plyusnin17, Plyusnin_2015}.
 
For all the scenarios, all the plasma parameters were kept constant in time, and a step in the electric field time evolution is introduced to observe the response of the system. For each case, the electric field value was chosen to have a moderate but observable runaway electron generation and to avoid the slide-away effect, where calculations of \textsc{Runaway Fluid} would be completely invalid. The magnetic field strength was set to be zero to omit the synchrotron radiation losses on the distribution function by \textsc{NORSE} and \textsc{DREAM}. Toroidicity effects were also omitted as \textsc{NORSE} cannot take them into account. The physical parameters for the different scenarios are listed in Table~\ref{params}.

\begin{figure*}[b]
\begin{center}
\includegraphics[height=15cm]{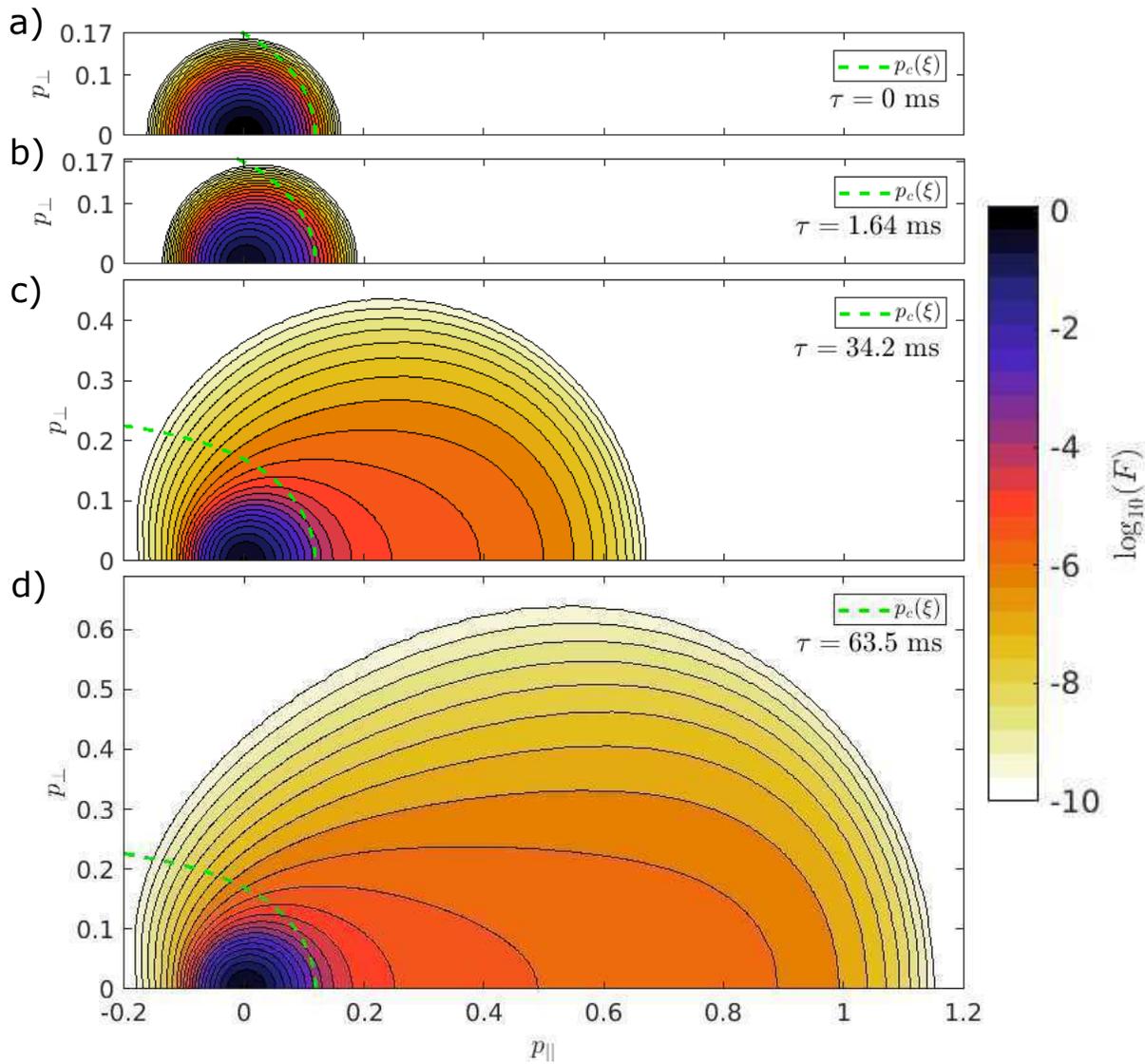}
\caption{The 2D distribution function throughout the simulation for the start--up case. The runaway boundary is marked with a dashed line.}
\label{distribution}
\end{center}
\end{figure*}

The evolution of the electron distribution function in 2D momentum space, as calculated by \textsc{NORSE}, is shown in Figure~\ref{distribution} for start--up scenario. The green dashed line shows the boundary of the runaway region calculated by \textsc{NORSE} based on an estimate by Smith et al.~\citep{Smith05}. In the first two plots, the shifting of the initial Maxwellian distribution into a Spitzer--like distribution dominates as the electric field is applied. During this time, the distribution function is still mostly isotropic as it shifts slightly towards the direction of the acceleration by the electric field. In later stages of the simulation, shown on the bottom two panels, we can see the formation of the high energy tail consisting of runaway electrons.

\subsection{Evolution of runaway generation rate}\label{sec:growth rate}

\begin{figure*}
\begin{center}
\includegraphics[width=1\linewidth]{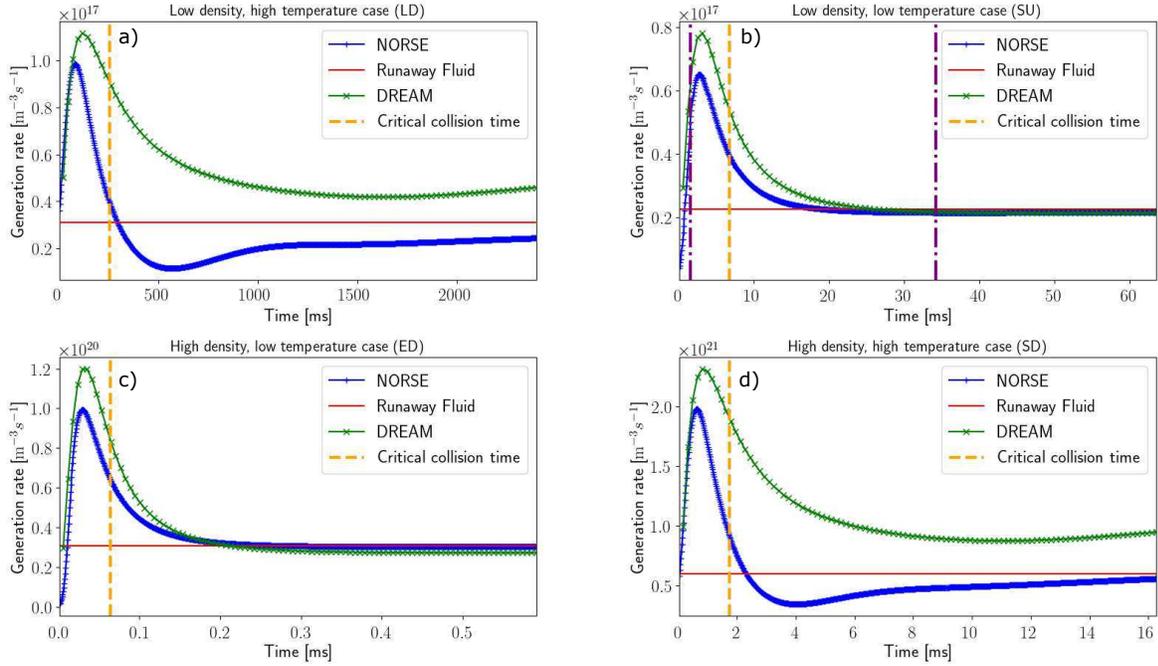}
\caption{The Dreicer generation rate calculated with \textsc{NORSE}, \textsc{Runaway Fluid} and \textsc{DREAM} in the four different cases. The collision time at the critical velocity is indicated on the plots with a vertical dashed line. The dot--dashed lines in the start--up (SU) case on plot (b) indicate the times where the distribution function is shown in Figure~\ref{distribution}.}
\label{gr}
\end{center}
\end{figure*}

The Dreicer generation rates from \textsc{NORSE} and \textsc{Runaway Fluid} are shown in Figure~\ref{gr} for all four scenarios. The plots look qualitatively similar, but the time scales are widely varying among the figures. A large peak in the generation rate from \textsc{NORSE} can be seen in all four cases. The peak might be caused by the slight but rapid shift and distortion of the whole Maxwellian into a Spitzer-like distribution, see Figure~\ref{distribution}, as the electric field is introduced. The overlap of the bulk population with the runaway region gives rise to the large but temporary peak in generation rate, which smooths out as runaway electrons are accelerated to higher energies. The generation later relaxes to the analytical generation rate value as the Maxwellian distribution develops a large runaway electron tail due to Dreicer generation. In the later stages, the runaway generation mechanism is the diffusion of electrons in the velocity space through the critical velocity boundary, as expected, and the kinetic generation rate closely follows the analytical calculations in the quasi--stationary limit~\citep{Dreicer60, Connor75}.

The same scenarios have also been run with the kinetic solver \textsc{DREAM}, which uses a test particle collision operator to calculate the Dreicer generation rate. The runaway boundary for the plotted generation rate was defined at $|p| = p_{\rm c} = 1/\sqrt{E/E_{\rm c} - 1}$, at which point collisional friction balances electric field acceleration for an electron with pitch xi=1~\citep{Connor75}. In the two low-temperature cases, SU and ED (plots (b) and (c)), the \textsc{DREAM} generation rate closely follows the analytical value in the later stages. The peak is also reproduced with \textsc{DREAM} although it has a slightly higher maximum than \textsc{NORSE}, most likely due to the differing runaway region definitions. In cases LD and SD a significant runaway electron fraction is generated. In Figure~\ref{gr} plots (a) and (d), \textsc{DREAM} slightly overestimates the generation rate even for later times, which is due to this decrease in bulk electron density. The fall after the peak does not reach the analytical value, and also a steady--state is not reached in the simulated time intervals. The interaction of runaway electrons with the bulk population could compensate for the increased runaway electron generation due to the reduced bulk density, but in DREAM this is not calculated leading to the increased generation rate. In these two cases, the \textsc{NORSE} generation rate undershoots the analytical value and approaches it from below. NORSE takes the runaway-bulk interaction into account which explains the lower generation rate compared to DREAM.

It can be concluded that there is a good qualitative agreement between \textsc{NORSE} and \textsc{DREAM}, and the time scale of the evolution of the initial peak is similar. A more thorough comparison of the two codes is out of scope for the current paper, as the observed differences do not affect the conclusions.

\subsection{Evolution of runaway density}\label{sec:runaway density}

\begin{figure*}
\begin{center}
\includegraphics[width=1\linewidth]{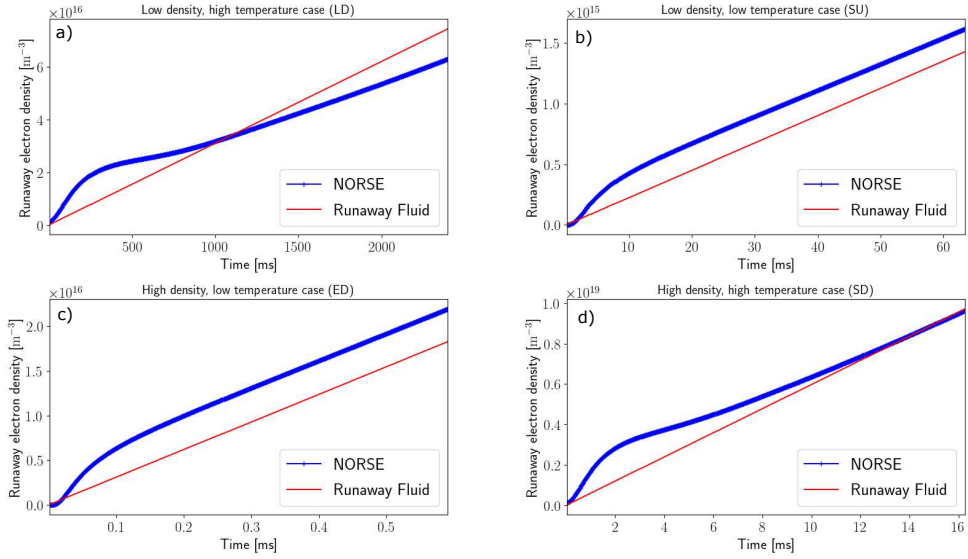}
\caption{The runaway electron density as a function of time calculated by \textsc{NORSE} and \textsc{Runaway Fluid} in the four scenarios.}
\label{density}
\end{center}
\end{figure*}

The runaway electron densities were also plotted as functions of time in all cases, as shown in Figure~\ref{density}. The generation rate peak observed in the kinetic calculations causes a constant shift in the runaway density between \textsc{NORSE} and \textsc{Runaway Fluid} for cases SU and ED (see plots (b) and (c)) where no undershoot was observed. In LD and SD cases (plots (a) and (d)), where the generation rate temporarily fell below the analytical generation rate before recovering to the asymptotic value, the runaway electron density from \textsc{NORSE} is slightly lower than the density from \textsc{Runaway Fluid} on long time scales, but the difference is not significant in any of the cases. The relative difference is smaller for more extended time scales in all scenarios. 

\subsection{Relevant time scales}\label{sec:time}

The characteristic time scale governing the Dreicer generation rate is related to the electron--electron collision time at the critical velocity. The critical velocity of runaway electron generation only depends on the electric field normalized to the critical field as
\begin{equation}
v_c=\frac{c}{\sqrt{\frac{E}{E_c}}},
\end{equation}
where $E_c$ is the critical electric field given by Equation~(\ref{Ec}). The velocity dependent electron-electron collision time at this velocity can be calculated as
\begin{equation}
\tau_{ee}=\frac{4 \pi \varepsilon_0^2 {m_e}^2v_c^3} {e^4 n_e \ln\Lambda}
\end{equation}
This quantity depends on three plasma parameters, namely the electron density, the temperature and the electric field. The density dependence appears explicitly and through the Coulomb logarithm in the denominator as well as through the critical field in the critical velocity. The temperature dependence comes solely from the Coulomb logarithm, and hence it has a weaker effect than the density or electric field. The strength of the electric field relative to the critical electric field will influence the collision time through the critical velocity. The explicit dependence of the collision time on these physical parameters is
\begin{equation}
\tau_{ee}=\frac{4 \pi \varepsilon_0^2 {m_e}^2c^3} {e^4 n_e \ln\Lambda \, \left(\frac{E}{E_c}\right)^{3/2}}.
\end{equation}
The critical field is dependent on density and temperature as shown in Equation~(\ref{Ec}) with the temperature dependence is again hidden in the Coulomb logarithm. Substitution to the collision time yields
\begin{equation}
\tau_{ee}(n, T, E)=\sqrt{\frac{m_e}{4\pi\varepsilon_0^2}} \frac{n_e^{1/2}\left(\ln\Lambda(n, T)\right)^{1/2}}{E^{3/2}},
\end{equation}
where the explicit dependence on the plasma parameters is expressed.

The relaxation of the generation rate peak happens on different absolute time scales for the different scenarios. The simulation time interval was chosen to achieve the steady--state solution from \textsc{NORSE} in all cases. The time to converge to the steady--state solutions is governed by the distortion of the distribution function at the boundary of the runaway region, which is driven by the collisions of the electrons. The time scale of the relaxation was found to be related to the collision time at the critical velocity of runaway electrons, and it was indicated in Figure~\ref{gr} with a vertical red line. It can be seen that the peak of the generation rate is already relaxing at the critical velocity collision time. On longer time scales, the kinetic and \textsc{Runaway Fluid} generation rates show good agreement with differences only caused by the varying definitions of the runaway region.

\begin{figure*}
\begin{center}
\includegraphics[width=1\linewidth]{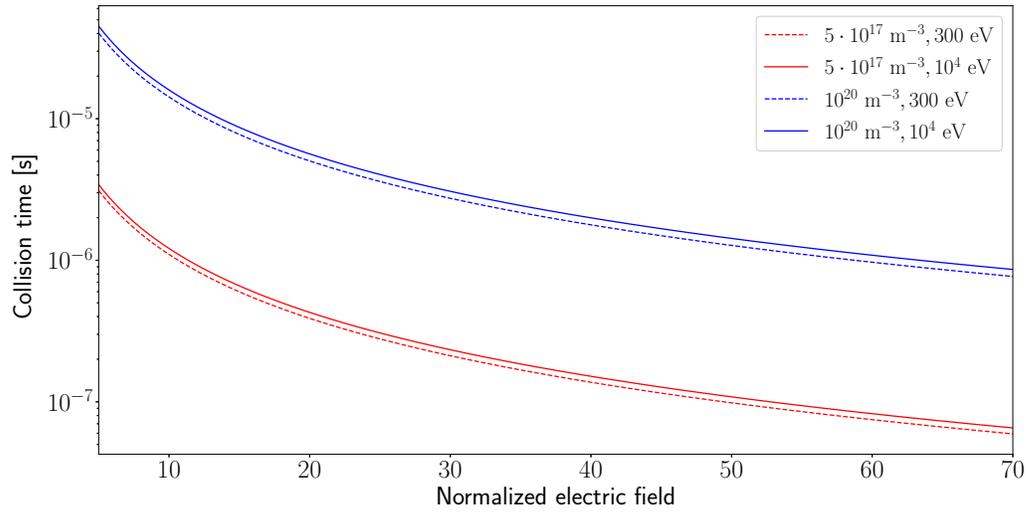}
\caption{The dependence of the collision time at the critical velocity on the normalized electric field.}
\label{tau}
\end{center}
\end{figure*}

The collision time as a function of the normalized electric field is shown in Figure~\ref{tau}. Cases with identical densities are indicated with the same colors. The low density scenarios are plotted in red while the high density in blue. Dashed lines are used for low temperatures, while solid lines for high temperatures. The temperature has only a small effect on the collision time and the main dependence is on density and electric field. 

\begin{figure*}
\begin{center}
\includegraphics[width=1\linewidth]{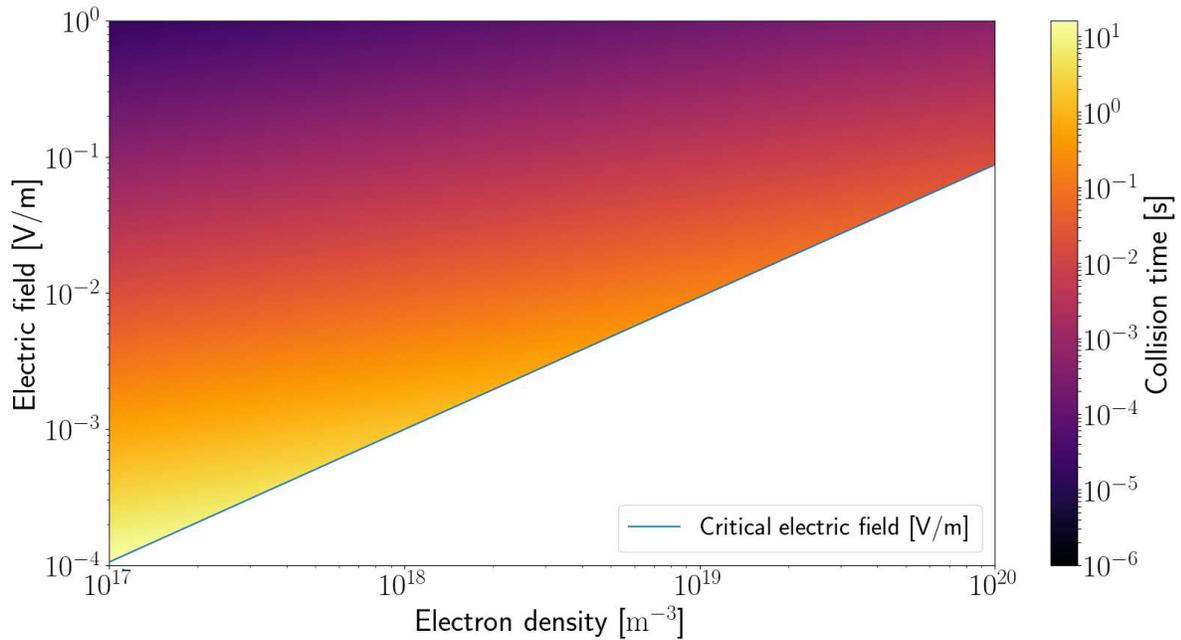}
\caption{The dependence of the collision time at the critical velocity on the density and the electric field.}
\label{tau2d}
\end{center}
\end{figure*}

In Figure~\ref{tau2d} the collision time is plotted as a function of density and electric field with constant temperature. The temperature was chosen to be $10~\;\rm keV$, although it does not significantly affect the dependence only the magnitude: the magnitude of the collision time is changed by a mere factor of about $1.2$ for $300 \;\rm eV$. The critical electric field is plotted in blue. The collision time covers about seven orders of magnitude on the parameter space, and it is not shown below the critical electric field where the critical velocity is not defined. Observe that for the extremely low density cases this collision time extends up to the several seconds region.

The ratio of the time scale of the simulated phenomenon related to the collision time at the critical velocity robustly indicates whether kinetic modelling is required. If the time scale is comparable or smaller than the collision time then the reduced kinetic modelling is not sufficiently accurate to simulate runaway electron production. In these cases, the slight but rapid shift of the distribution function generates a significantly larger runaway population relative to the quasi-stationary analytical estimates. If the phenomenon in question is quasi-stationary on a  significantly longer time scale than the critical velocity collision time, the generation rates from the full kinetic and the reduced kinetic models show good agreement, and the latter can be used with good accuracy.

\section{Discussion\label{sec:discussion}}

The usage of the step function to investigate the response of a system to rapid change is a common choice in system theory for linear and non-linear systems~\citep{Feucht90}. This motivated the introduction of a step in the electric field strength to study the time response of the electron distribution. It is physically not realistic to have an instantaneous change in any parameter, but the step functions has been shown to reveal the dynamical properties of the underlying system. The system response to a rapid change can be characterized by different parameters, such as the rise time to reach the new steady--state value, the overshoot of the output quantity, or the settling time after the overshoot~\citep{Feucht90}.

A neural network has recently been used to fit the steady--state  Dreicer generation rate~\citep{Hesslow19} to kinetic simulations. The similarity of our results with the linear system step response in control theory~\citep{Feucht90} suggests that a linear system fitted by a neural network like the one by Hesslow et al.~\citep{Hesslow19}, could potentially be used to estimate the response of Dreicer generation rate to rapid change in plasma parameters.

Another transient runaway electron generation mechanism is the so--called hot--tail generation of runaway electrons which typically occurs during tokamak disruptions~\citep{Smith08, Aleynikov17, Breizman19, Martin_Solis17}, when the energetic tail of the Maxwellian distribution can end up in the runaway region. This is possible when the time scale of the plasma cooling is shorter than the electron--electron collision time, as the fast particles will not have time to thermalize before the critical boundary for runaway electrons in the momentum space is lowered due to the increasing electric field in the current quench~\citep{Smith08}. The cooling rate required for hot--tail generation by Smith and Verwichte~\citep{Smith08} is also related to the fast electron collision time. Such fast cooling of the background plasma is possible during the thermal quench phase of disruptions, and it has been indicated~\citep{Aleynikov17, Breizman19, Martin_Solis17} that in ITER disruptions the hot--tail generation will dominate over the Dreicer generation. However, in ASDEX Upgrade and JET disruptions, the Dreicer generation can be comparable to the hot--tail generation~\citep{Bjork2021, linder2021}, hence accurate simulation of the Dreicer generation rate is relevant and necessary in simulations of disruptions in current-day devices. In the other studied cases, the runaway seed is expected to be generated mostly by the Dreicer mechanism. In the start--up phase and the low density discharge case Dreicer generation can create sufficient seed for the avalanche generation, and thus, the accurate modelling of runaway generation is crucial.

The avalanche generation of runaway electrons plays a vital role when a seed of runaway electrons has already been generated by primary generation processes, such as Dreicer or hot--tail generation. The avalanche effect can amplify the runaway electron beam significantly. In a simple simulation where prescribed plasma parameters are used, a discrepancy in the runaway seed calculated by the kinetic and reduced kinetic models can be magnified exponentially in the final runaway electron density by the avalanche mechanism. Self--consistent evolution of the plasma parameters, on the other hand, would create negative feedback from the runaway current on the induced electric field, resulting in the reduction of the sensitivity of the final runaway electron density on the initial seed~\citep{Breizman19}.

\section{Conclusion}\label{sec:conclusion}

We have investigated three different models for Dreicer generation: a reduced kinetic, a linear kinetic and a non-linear kinetic. We have studied four different scenarios meant to represent the full experimentally relevant range. \textsc{Runaway Fluid} uses analytical formulas to calculate the runaway generation rates. \textsc{NORSE} and \textsc{DREAM} were used to calculate the evolution of the electron distribution function with non-linear and linearized collision operators. The Runaway Electron Test Workflow was developed in the European Integrated Modelling framework where the different codes were used in parallel with identical input parameters and output data structures, enabling the comparison of the results with relative ease. 

We investigated the response of the electron distribution function to a jump in the electric field with the aim to study the relevant time scales and kinetic features of the Dreicer runaway generation. We found that the kinetic models show a large initial peak in the runaway electron generation rate. We observed a rapid initial shift and distortion in the Maxwellian electron distribution, which might cause the peak as the electric field is introduced. The peak in the generation rate cannot be captured by reduced kinetic models using the traditional quasi-stationary analytical generation rates. Moreover, it is relaxed as the anisotropization of the distribution function develops. The generation rate calculated from the kinetic solvers converges to the analytical value, and the convergence is driven by the collision of electrons on a time scale related to the collision time at the critical velocity for runaway electrons. This kinetic effect in the generation rate creates a constant shift of runaway electron density calculated by the different models. This shift is not significant if only Dreicer generation is considered, and the relative difference is reduced with long timescales. 

Our main conclusion is that on time scales comparable or shorter than the electron--electron collision time calculated at the critical velocity for runaway electron generation, time-dependent kinetic modelling is required to capture the relevant physical processes during Dreicer generation of runaway electrons. We find that transient effects can occur on time scales comparable to this collision time, which could be important to consider in self--consistent simulations of tokamak start--up~\citep{Sharma88, deVries20}, disruptions~\citep{Papp13} or low density discharges~\citep{Plyusnin_2015, Plyusnin17}. The temperature dependence of the time scale is weak, but for low density and low electric field cases, it can reach the several seconds range. Calculation of the collision time can ensure the use of optimal runaway electron modelling tools in complex self--consistent simulations of start--up and disruption scenarios.

    \section*{Acknowledgments} 
The authors are grateful to G.~Por, O.~Linder and G.~Papp for fruitful discussions. This work has been carried out within the framework of the EUROfusion Consortium and has received funding from the Euratom research and training programme 2014--2018 and 2019--2020 under grant agreement No 633053. The views and opinions expressed herein do not necessarily reflect those of the European Commission. G.~I.~Pokol and S.~Olasz acknowledge the support of the National Research, Development and Innovation Office (NKFIH) Grant FK132134. 

\bibliographystyle{unsrt}

\bibliography{main}

\begin{thebibliography}{10}

\bibitem{Boozer17}
A.~H. Boozer.
\newblock Runaway electrons and {ITER}.
\newblock {\em Nuclear Fusion}, 57(5):056018, 2017.

\bibitem{Lehnen15}
M.~Lehnen, K.~Aleynikova, P.B. Aleynikov, D.~Campbell, P.~Drewelow, N.W.
  Eidietis, Y.~Gasparyan, R.S. Granetz, Y.~Gribov, N.~Hartmann, E.~Hollmann,
  V.~Izzo, S.~Jachmich, K.~Sangjin, M.~Kočan, H.~R. Koslowski, D.~Kovalenko,
  U.~Kruezi, A.~Loarte, and P.~De~Vries.
\newblock Disruptions in {ITER} and strategies for their control and
  mitigation.
\newblock {\em Journal of Nuclear Materials}, 463:39--48, 2015.

\bibitem{Zeng17}
L.~Zeng, Z.Y. Chen, Y.B. Dong, H.R. Koslowski, Y.~Liang, Y.P. Zhang, H.D.
  Zhuang, D.W. Huang, and X.~Gao.
\newblock Runaway electron generation during disruptions in the {J-{TEXT}}
  tokamak.
\newblock {\em Nuclear Fusion}, 57(4):046001, 2017.

\bibitem{Sharma88}
A.S. Sharma and R.~Jayakumar.
\newblock Runaway electrons during tokamak startup.
\newblock {\em Nuclear Fusion}, 28(3):491--498, 1988.

\bibitem{deVries20}
P.~C. de~Vries, Y.~Gribov, R.~Martin-Solis, A.~B. Mineev, J.~Sinha, A.~C.~C.
  Sips, V.~Kiptily, and A.~Loarte.
\newblock Analysis of runaway electron discharge formation during {Joint
  European Torus} plasma start-up.
\newblock {\em Plasma Physics and Controlled Fusion}, 62(12):125014, 2020.

\bibitem{Plyusnin06}
V.V. Plyusnin, V.~Riccardo, R.~Jaspers, B.~Alper, V.G. Kiptily, J.~Mlynar,
  S.~Popovichev, E.~de~La~Luna, F.~Andersson, and JET~EFDA contributors.
\newblock Study of runaway electron generation during major disruptions in
  {JET}.
\newblock {\em Nuclear Fusion}, 46(2):277--284, 2006.

\bibitem{Plyusnin17}
V.V. Plyusnin, C.~Reux, V.G. Kiptily, G.~Pautasso, J.~Decker, G.~Papp,
  A.~Kallenbach, V.~Weinzettl, J.~Mlynar, S.~Coda, V.~Riccardo, P.~Lomas,
  S.~Jachmich, A.E. Shevelev, B.~Alper, E.~Khilkevitch, Y.~Martin, R.~Dux,
  C.~Fuchs, B.~Duval, M.~Brix, G.~Tardini, M.~Maraschek, W.~Treutterer,
  L.~Giannone, A.~Mlynek, O.~Ficker, P.~Martin, S.~Gerasimov, S.~Potzel,
  R.~Paprok, P.~J. McCarthy, M.~Imrisek, A.~Boboc, K.~Lackner, A.~Fernandes,
  J.~Havlicek, L.~Giacomelli, M.~Vlainic, M.~Nocente, and U.~Kruezi.
\newblock Comparison of runaway electron generation parameters in small,
  medium-sized and large tokamaks{\textemdash}a survey of experiments in
  {COMPASS}, {TCV}, {ASDEX}-upgrade and {JET}.
\newblock {\em Nuclear Fusion}, 58(1):016014, 2017.

\bibitem{Dreicer60}
H.~Dreicer.
\newblock Electron and ion runaway in a fully ionized gas. ii.
\newblock {\em Phys. Rev.}, 117:329--342, 1960.

\bibitem{Connor75}
J.W. Connor and R.J. Hastie.
\newblock Relativistic limitations on runaway electrons.
\newblock {\em Nuclear Fusion}, 15(3):415--424, 1975.

\bibitem{Pokol19}
G.~I. Pokol, S.~Olasz, B.~Erdos, G.~Papp, M.~Aradi, M.~Hoppe, T.~Johnson,
  J.~Ferreira, D.~Coster, Y.~Peysson, J.~Decker, P.~Strand, D.~Yadikin,
  D.~Kalupin, and the EUROfusion-IM~Team.
\newblock Runaway electron modelling in the self-consistent core {European
  Transport Simulator}.
\newblock {\em Nuclear Fusion}, 59, 2019.

\bibitem{Imbeaux15}
F.~Imbeaux, S.D. Pinches, J.B. Lister, Y.~Buravand, T.~Casper, B.~Duval,
  B.~Guillerminet, M.~Hosokawa, W.~Houlberg, P.~Huynh, S.H. Kim, G.~Manduchi,
  M.~Owsiak, B.~Palak, M.~Plociennik, G.~Rouault, O.~Sauter, and P.~Strand.
\newblock Design and first applications of the {ITER} integrated modelling {\&}
  analysis suite.
\newblock {\em Nuclear Fusion}, 55(12):123006, 2015.

\bibitem{Papp13}
G.~Papp, T.~Fülöp, T.~Fehér, P.C. de~Vries, V.~Riccardo, C.~Reux, M.~Lehnen,
  V.~Kiptily, V.V. Plyusnin, B.~Alper, and JET~EFDA contributors.
\newblock The effect of {ITER}-like wall on runaway electron generation in
  {JET}.
\newblock {\em Nuclear Fusion}, 53, 2013.

\bibitem{Bandaru19}
V.~Bandaru, M.~Hoelzl, F.~J. Artola, G.~Papp, and G.~T.~A. Huijsmans.
\newblock Simulating the nonlinear interaction of relativistic electrons and
  tokamak plasma instabilities: Implementation and validation of a fluid model.
\newblock {\em Phys. Rev. E}, 99:063317, 2019.

\bibitem{Linder20}
O.~Linder, E.~Fable, F.~Jenko, G.~Papp, and G.~Pautasso.
\newblock Self-consistent modeling of runaway electron generation in massive
  gas injection scenarios in {ASDEX} {Upgrade}.
\newblock {\em Nuclear Fusion}, 60(9):096031, 2020.

\bibitem{Stahl16}
A.~Stahl, M.~Landreman, O.~Embréus, and T.~Fülöp.
\newblock {NORSE}: A solver for the relativistic non-linear fokker–planck
  equation for electrons in a homogeneous plasma.
\newblock {\em Computer Physics Communications}, 212, 2016.

\bibitem{DREAM}
M.~Hoppe, O.~Embreus, and T.~F\"ul\"op.
\newblock {DREAM}: a fluid-kinetic framework for tokamak disruption runaway
  electron simulations.
\newblock {\em In preparation}, 2021.

\bibitem{Landreman13}
M.~Landreman, A.~Stahl, and T.~Fülöp.
\newblock Numerical calculation of the runaway electron distribution function
  and associated synchrotron emission.
\newblock {\em Computer Physics Communications}, 185(3):847 -- 855, 2014.

\bibitem{Stahl2016}
A.~Stahl, O.~Embr{\'{e}}us, G.~Papp, M.~Landreman, and T.~Fülöp.
\newblock Kinetic modelling of runaway electrons in dynamic scenarios.
\newblock {\em Nuclear Fusion}, 56(11):112009, 2016.

\bibitem{del-Castillo-Negrete18}
D.~del Castillo-Negrete, L.~Carbajal, D.~Spong, and V.~Izzo.
\newblock Numerical simulation of runaway electrons: {3-D} effects on
  synchrotron radiation and impurity-based runaway current dissipation.
\newblock {\em Physics of Plasmas}, 25(5):056104, 2018.

\bibitem{Beidler20}
M.~T. Beidler, D.~del Castillo-Negrete, L.~R. Baylor, D.~Shiraki, and D.~A.
  Spong.
\newblock Spatially dependent modeling and simulation of runaway electron
  mitigation in {DIII-D}.
\newblock {\em Physics of Plasmas}, 27(11):112507, 2020.

\bibitem{Peysson14}
Y.~Peysson and J.~Decker.
\newblock Numerical simulations of the radio-frequency–driven toroidal
  current in tokamaks.
\newblock {\em Fusion Science and Technology}, 65(1):22--42, 2014.

\bibitem{Decker16}
J.~Decker, E.~Hirvijoki, O.~Embréus, Y.~Peysson, A.~Stahl, I.~Pusztai, and
  T.~Fülöp.
\newblock Numerical characterization of bump formation in the runaway electron
  tail.
\newblock {\em Plasma Physics and Controlled Fusion}, 58:025016, 2016.

\bibitem{Feucht90}
Dennis~L. Feucht.
\newblock {\em Handbook of Analog Circuit Design}.
\newblock Academic Press, 1990.

\bibitem{Falchetto14}
G.L. Falchetto, D.~Coster, R.~Coelho, B.D. Scott, L.~Figini, D.~Kalupin,
  E.~Nardon, S.~Nowak, L.L. Alves, J.F. Artaud, V.~Basiuk, Jo{\~{a}}o~P.S.
  Bizarro, C.~Boulbe, A.~Dinklage, D.~Farina, B.~Faugeras, J.~Ferreira,
  A.~Figueiredo, Ph. Huynh, F.~Imbeaux, I.~Ivanova-Stanik, T.~Jonsson, H.-J.
  Klingshirn, C.~Konz, A.~Kus, N.B. Marushchenko, G.~Pereverzev, M.~Owsiak,
  E.~Poli, Y.~Peysson, R.~Reimer, J.~Signoret, O.~Sauter, R.~Stankiewicz,
  P.~Strand, I.~Voitsekhovitch, E.~Westerhof, T.~Zok, W.~Zwingmann, and and.
\newblock The {European Integrated Tokamak Modelling} ({ITM}) effort:
  achievements and first physics results.
\newblock {\em Nuclear Fusion}, 54(4):043018, 2014.

\bibitem{Smith08}
H.~M. Smith and E.~Verwichte.
\newblock Hot tail runaway electron generation in tokamak disruptions.
\newblock {\em Physics of Plasmas}, 15(7):072502, 2008.

\bibitem{Bjork20}
K.~Insulander~Björk, G.~Papp, O.~Embreus, L.~Hesslow, T.~Fülöp,
  O.~Vallhagen, A.~Lier, G.~Pautasso, A.~Bock, ASDEX~Upgrade Team, and
  Eurofusion~MST1 Team.
\newblock Kinetic modelling of runaway electron generation in argon-induced
  disruptions in {ASDEX Upgrade}.
\newblock {\em Journal of Plasma Physics}, 86(4), 2020.

\bibitem{Aleynikov17}
P.~Aleynikov and B.~N. Breizman.
\newblock Generation of runaway electrons during the thermal quench in
  tokamaks.
\newblock {\em Nuclear Fusion}, 57(4):046009, 2017.

\bibitem{Helander04}
P.~Helander, H.~Smith, T.~Fülöp, and L.-G. Eriksson.
\newblock Electron kinetics in a cooling plasma.
\newblock {\em Physics of Plasmas}, 11(12):5704--5709, 2004.

\bibitem{Rosenbluth97}
M.N. Rosenbluth and S.V. Putvinski.
\newblock Theory for avalanche of runaway electrons in tokamaks.
\newblock {\em Nuclear Fusion}, 37(10):1355--1362, 1997.

\bibitem{Huba13}
J.~D. Huba.
\newblock {\em {NRL Plasma Formulary Supported by The Office of Naval
  Research}}.
\newblock Naval Research Laboratory, Washington, DC, 2013.

\bibitem{Plyusnin_2015}
V.V. Plyusnin, V.G. Kiptily, A.E. Shevelev, E.M. Khilkevitch, M.~Brix,
  S.~Gerasimov, G.~F. Matthews, and JET contributors.
\newblock Parameters and stability of runaway electron dominating discharge in
  jet with iter-like wall.
\newblock In R.~Bingham, W.~Suttrop, S.~Atzeni, R.~Foest, K.~McClements,
  B.~Gonçalves, C.~Silva, and R.~Coelho, editors, {\em 42nd EPS Conference on
  Plasma Physics, EPS 2015}. European Physical Society {EPS}, 2015.

\bibitem{Smith05}
H.~Smith, P.~Helander, L.-G. Eriksson, and T.~Fülöp.
\newblock Runaway electron generation in a cooling plasma.
\newblock {\em Physics of Plasmas}, 12(12):122505, 2005.

\bibitem{Hesslow19}
L.~Hesslow, L.~Unnerfelt, O.~Vallhagen, O.~Embréus, M.~Hoppe, G.~Papp, and
  T.~Fülöp.
\newblock Evaluation of the {Dreicer} runaway growth rate in the presence of
  {high-Z} impurities using a neural network.
\newblock {\em Journal of Plasma Physics}, 85, 2019.

\bibitem{Breizman19}
B.~N. Breizman, P.~Aleynikov, E.~M. Hollmann, and M.~Lehnen.
\newblock Physics of runaway electrons in tokamaks.
\newblock {\em Nuclear Fusion}, 59(8):083001, 2019.

\bibitem{Martin_Solis17}
J.R. Mart{\'{\i}}n-Sol{\'{\i}}s, A.~Loarte, and M.~Lehnen.
\newblock Formation and termination of runaway beams in {ITER} disruptions.
\newblock {\em Nuclear Fusion}, 57(6):066025, 2017.

\bibitem{Bjork2021}
K.~Insulander Björk, O.~Vallhagen, G.~Papp, C.~Reux, O.~Embreus, E.~Rachlew,
  T.~Fülöp, the ASDEX Upgrade~Team, JET contributors, and the EUROfusion
  MST1~Team.
\newblock Modelling of runaway electron dynamics during argon-induced
  disruptions in {ASDEX Upgrade} and {JET}.
\newblock 2021.

\bibitem{linder2021}
O.~Linder, G.~Papp, E.~Fable, F.~Jenko, G.~Pautasso, the ASDEX Upgrade~Team,
  and the EUROfusion MST1~Team.
\newblock Electron runaway in {ASDEX Upgrade} experiments of varying core
  temperature.
\newblock {\em Submitted to the Journal of Plasma Physics}, 2021.

\end{thebibliography}
\end{document}